\documentclass[%
reprint,
superscriptaddress,
amsmath,amssymb,
aps,
]{revtex4-2}
\usepackage{chemformula} 
\usepackage[version=3]{mhchem}
\usepackage{caption}
\usepackage{siunitx}
\usepackage{subcaption}
\usepackage{graphicx}
\usepackage{dcolumn}
\usepackage{bm}
\usepackage{upgreek}
\usepackage{todonotes}
\usepackage[]{xcolor}

\begin{document}
	
	\preprint{APS/123-QED}

\title{
	Light-induced atomic motion in ionic crystals}

\author{Jacob C. Warming}
\affiliation{%
	Department of Physics \& Astronomy, Aarhus University, Ny Munkegade 120, 8000 Aarhus, Denmark 
}
\author{ Simon P. S. Jessen}
\affiliation{%
	Department of Physics \& Astronomy, Aarhus University, Ny Munkegade 120, 8000 Aarhus, Denmark 
}
\author{Emma A. Husted}
\affiliation{%
	Department of Physics \& Astronomy, Aarhus University, Ny Munkegade 120, 8000 Aarhus, Denmark 
}
\author{Jan Thøgersen}
\affiliation{%
	Department of Chemistry, Aarhus University, Langelandsgade 140, 8000 Aarhus, Denmark
}
\author{Stefan Gundacker}
\affiliation{%
	Marietta Blau Institute for Particle Physics, Austrian Academy of Sciences, Nikolsdorfer Gasse 18, 1050 Vienna, Austria
}
\author{ Brian Julsgaard}
\affiliation{%
	Department of Physics \& Astronomy, Aarhus University, Ny Munkegade 120, 8000 Aarhus, Denmark 
}
\author{ Peter Balling}
\affiliation{%
	Department of Physics \& Astronomy, Aarhus University, Ny Munkegade 120, 8000 Aarhus, Denmark 
}
\author{Rosana M. Turtos}
\email[Email: ]{ro.turtos@phys.au.dk}
\affiliation{%
	Department of Physics \& Astronomy, Aarhus University, Ny Munkegade 120, 8000 Aarhus, Denmark 
}


\definecolor{myred}{RGB}{242,126,149}
\definecolor{myblue}{RGB}{142,126,249}
\definecolor{mygreen}{RGB}{142,226,119}

\begin{abstract}
Atomic motion in solids is conventionally driven by elastic collisions between ionizing particles and atoms, which transfer momentum and induce lattice displacements. In this work, we demonstrate a different mechanism for atomic displacement based on optical excitation of scintillating ionic crystals. 	Ionic crystals are unique systems because of the closed-shell electronic configuration of their constituent ions. In these materials, excitation above the band gap generates a hole that strongly distorts the lattice, resulting in the formation of a self-trapped hole (STH). The STH is Coulomb-attracted to the electron, thereby forming a self-trapped exciton (STE). Here, we demonstrate that in \ce{BaF2} – one of the fastest scintillators – the STE structure promotes
the formation of long-lived electron and hole traps that persist in the lattice at room temperature. Such trapped
electron-hole pairs occupy vacancy-interstitial fluorine pair positions, and can be
created indiscernibly using optical or ionizing radiation excitation, as long as the STH is formed. 
Further, we demonstrate that it is possible to control the defect evolution with light. Selective optical stimulation of the trapped electrons or holes enables the regeneration of the STE at later times. This light-controlled defect engineering allows us to increase the yield of the STE signal appearing as optically stimulated luminescence (OSL) and to image the spatial distribution of the initial energy deposition, holding strong potential for ionizing-radiation detection. These findings provide a common framework underlying scintillation and OSL in ionic crystals of the fluorite structure, allowing for optical manipulation of atomic vacancies-interstitial pairs in similar systems.


\end{abstract}

\maketitle

\flushbottom


\section{Introduction}

Scintillators are materials that exhibit a strong optical response to ionizing radiation and serve as the foundation for room-temperature ionizing-radiation spectroscopy. They support modern society in a broad palette of applications ranging from medical imaging and radiation dosimetry\cite{Lecoq2016,Dujardin2018} to nuclear monitoring, environmental sensing, and particle-physics research\cite{Lecoq2020,Baiocco2025}.
Here, we explore a complementary luminescence pathway, known as optically stimulated luminescence (OSL)\cite{Yukihara2011}, which effectively records the scintillation process.
Acting as a latent optical memory, it enables the storage and subsequent retrieval of spatial information associated with ionizing events, opening new opportunities for radiation detection spectroscopy that extend beyond the transient nature of the scintillation signal.

\begin{figure*}[ht]
\centering
\includegraphics[width=0.95\linewidth]{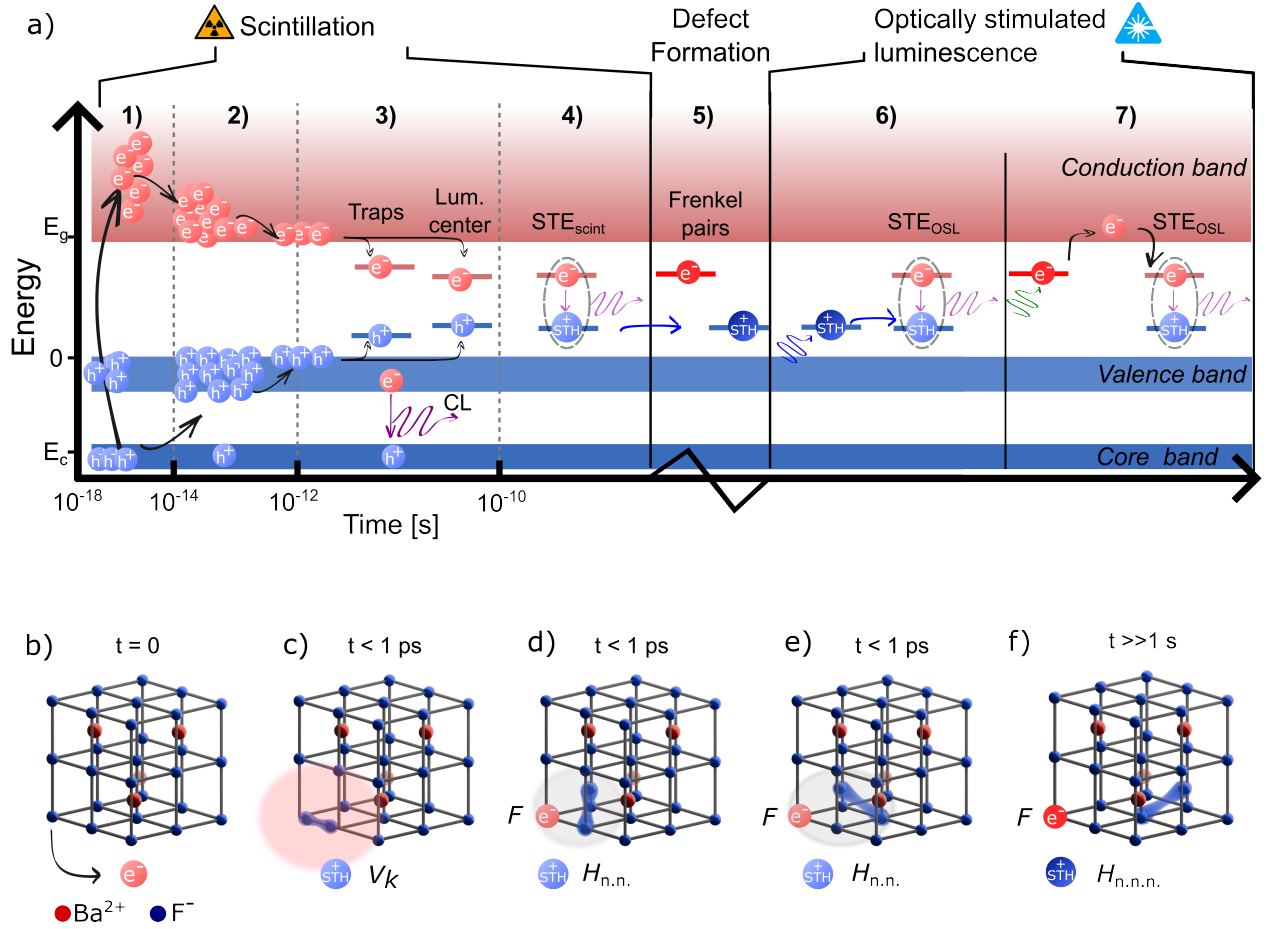}
\caption{\textbf{Luminescence and trapping mechanisms involved in the formation and decay of self-trapped excitons in \ce{BaF2}, both promptly during scintillation or in a delayed manner during optically stimulated luminescence (OSL)}. \textbf{a)} General scheme of relaxation of electronic excitation characterizing the scintillation multistage process in solid-state crystalline materials (1-3), with emphasis on intrinsic materials where the luminescence center is a self-trapped-exciton (STE) (4). The scheme has been extended to include the OSL mechanisms that are  open for examination in this work, and a time axis is used to depict localized and delocalized states in a band diagram. 
Free carriers in the form of electron-hole pairs are denoted by \ce{e-} and \ce{h+} symbols, while localized excitation in the form of a self-trapped hole is denoted as STH+. Notice that localized electrons in states within the bandgap states are also denoted with an \ce{e-} symbol to highlight the nature of the carrier. All non-wavy arrows are representative of non-radiative processes while wavy arrows indicate optical photon absorption or emission. \textbf{b-f)} Lattice configurations involved in the scintillation, defect formation, and OSL mechanisms specific to \ce{BaF2}. \textbf{b)} Cubic fluorite structure of the alkaline-earth fluoride family, where electron excitation across the bandgap is denoted by removing an electron from a fluorine ion. \textbf{c)} On-center STE = \ce{e-}~+~\emph{V$_K$}. \textbf{d)} Off-center STE = \emph{F}~+~\emph{H} of type II \cite{Lindner2001}. \textbf{e)} Off-center STE =  \emph{F}~+~\emph{H} of type III \cite{Lindner2001}. \textbf{f)} Lattice configuration for spatially separated \emph{F} and \emph{H} pairs. In this case, the \emph{H}~center is shown and denoted in the next-near-neighbor configuration \emph{H$_{n.n.n.}$}. }
\label{fig:1}
\end{figure*}

In a general framework, the optical photons appearing during scintillation are emitted following a multistage process 
which culminates with the radiative decay of a given optical transition\cite{Mikhail} and is illustrated in Fig.~\ref{fig:1}a. In crystalline systems, such a multistage process can be generalized in four sequential stages: (1) multiplication of electronic excitation, (2) thermalization of excited carriers, (3) transfer of electron-hole pairs to in-gap states, belonging to trapping or luminescence centers, and (4) radiative or non-radiative recombination of electron hole pairs.
When the crystal specifically has an ionic nature and the luminescence center is intrinsic, the optical transition responsible for scintillation in stage 4 will have the characteristics of a bound state between a self-trapped-hole (STH) and an electron, i.e., a self-trapped-exciton, denoted as STE$_\text{scint}$\cite{Williams1990}. After scintillation, some of the electron/hole pairs remain trapped in the crystal in metastable states and can later be optically stimulated, allowing for the trapped electron or hole to radiatively recombine with its counterpart particle. This optically stimulated luminescence (OSL) is very useful in advanced radiation dosimetry, however its origin 
remains obscure in materials where scintillation occurs via the STE pathway \cite{Evans1978,Yukihara2006}.

In this work, we elucidate the connection between the scintillation and OSL mechanisms using \ce{BaF2}. This is an interesting ionic crystal of the alkaline-earth fluoride family with a cubic fluorite crystalline structure, as shown in Fig.~\ref{fig:1}b, and stands out as one of the fastest bulk scintillators due to its cross-luminescence (CL) signal (see panel 3 of Fig.~\ref{fig:1}a). CL arises from optical transitions between valence-band electrons and core-band holes\cite{Aleksandrov1984}.
Besides this ultrafast response, scintillation in \ce{BaF2} is established to originate from the radiative recombination of an STE\cite{Williams1976}.
The STE is formed when a hole at the top of the valence band, i.e., essentially the 2p$^6$ orbitals of the fluoride ions, self-traps by forming a molecular-like configuration involving a neutral fluor and a nearby fluoride ion. This STH binds to a conduction-band electron, thereby forming the STE with an on-center structure denoted as \ce{e-}~+~\emph{V$_K$}, as shown in Fig.~\ref{fig:1}c, where the electron is delocalized. The system further relaxes to a lattice configuration where the STH rotates to create a fluor interstitial, i.e., an \emph{H}~center in the <111> direction (Fig.~\ref{fig:1}d-e), as confirmed by EPR measurements \cite{Beaumont1970, Hayes1974}. 
The electron localizes at the associated fluor vacancy forming the well-known \emph{F}~center\cite{Stoneham1968}. This is the common off-center STE structure characteristic of the fluorite crystal structure, denoted as \emph{F}~+~\emph{H}. 
Fig.~\ref{fig:1}d-e show the two lattice configurations that have been reported as most probable\cite{Lindner2001}, both being near-neighbor configurations of the \emph{H} center denoted as \emph{H$_{n.n.}$}.

Notice that the lattice configurations of the STE$_\text{scint}$ is equivalently a vacancy-interstitial fluorine pair in the near-neighbor position occupied by the electron and the STH. This STE is shown in stage 4 of the scintillation mechanism, decaying radiatively 
or non-radiatively. 
The non-radiative decay pathway creates spatially separated \emph{F} and \emph{H}~centers as illustrated in stage 5) and by the lattice configuration of Fig.~\ref{fig:1}f, where the \emph{H}~center is shown in an arbitrary next-near-neighbor position designated as \emph{H$_{n.n.n.}$}.

Here, we propose that those spatially-separated \emph{F} and \emph{H}~centers are the electron and hole traps that give rise to the OSL signal in \ce{BaF2}. 
We demonstrate that OSL is driven by the same STE emission observed during scintillation, although in a different lattice configuration. This forms the basis for processes 6) and 7) in the relaxation scheme of Fig.~\ref{fig:1}a and widens the range of applicability for the theory proposed by K. S. Song and R. T. Williams for alkali halide crystals\cite{Williams1978, Song1996_book, Song1995}, which attributes stable defect formation to the non-radiative decay of the STE. This mechanism has been experimentally verified in alkali halides \cite{Lushchik2000, Klinger1985} and was recently observed at room-temperature in copper-doped \ce{LiF} \cite{Nielsen2023}. 

We are further able to selectively stimulate the electron or hole traps and recreate the STE$_\text{OSL}$. This selective stimulation uses transport through the bands or molecular transport through the lattice as illustrated by stage 6) and 7) in Fig.~\ref{fig:1}a.
This allows us to increase the luminescence yield of the OSL signal for subsequent irradiations and readouts. Lastly, we demonstrate the ability to image the spatial distribution of initial energy deposition in such crystals.
These findings unify the excited-state framework for scintillation and OSL in fluorite-structure ionic crystals, enabling optical manipulation of vacancy–interstitial pairs in similar systems if the optical absorption of the traps remain spectrally separated.

\section{Results}
\begin{figure*}
    \centering
    \includegraphics[width=0.95\linewidth]{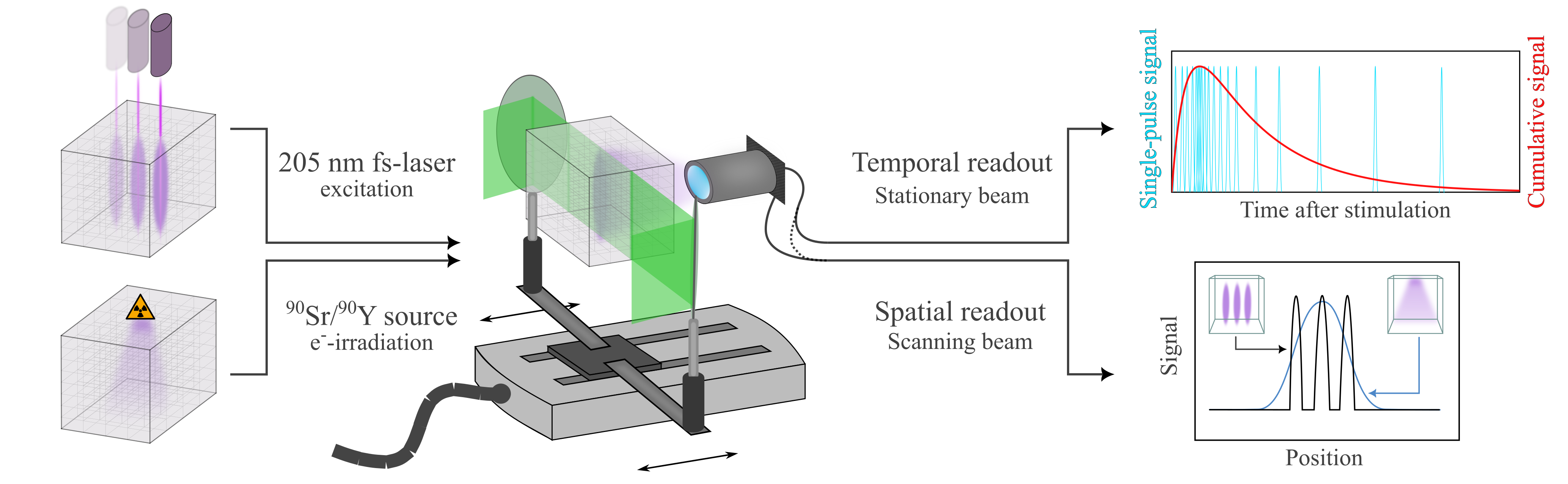}
    \caption{
    \textbf{Illustration of setup for high-sensitivity OSL readout.}
    The crystal is first excited with fs-laser pulses or irradiated by high-energy electrons. It is then placed in the setup where a lens objective couples light into a silicon photomultiplier (SiPM) photodetector. The OSL readout setup can be operated in two configurations. 1: The mirrors can be positioned at a point of irradiation and, using lower laser power as stimulation, the single-photon output from the high-frequency board of the SiPM produces a decay curve of the OSL by combining photon arrival times from several laser pulses. This is exemplified by the high-frequency response of one pulse in blue, and the resulting decay curve in red. 2: The stage moves the mirrors to scan the collimated laser sheet along the crystal, giving spatial information about the trapped charges.}
    \label{fig:2}
\end{figure*}

\subsection{Experimental overview}
Experiments were performed with \SI{1}{\centi\meter\cubed} \ce{BaF2} single crystals of different purity:~the purest optical-grade crystals from Hellma Materials and scintillator-optimized crystals from Epic Crystal.
Several types of experiments were performed to study the luminescence- and trapping-mechanisms in \ce{BaF2}, using conventional absorption and emission spectroscopy, time-resolved absorption, and OSL measurements under continuous-wave stimulation alongside a specially-tailored OSL setup. This setup is illustrated in Fig.~\ref{fig:2} and described in Methods. In Supplementary S1, we compare the efficacy of this setup with conventional temporal methods.

\subsection{Emission and decay kinetics}
\begin{figure*}
    \centering
    \includegraphics[width=0.9\linewidth]{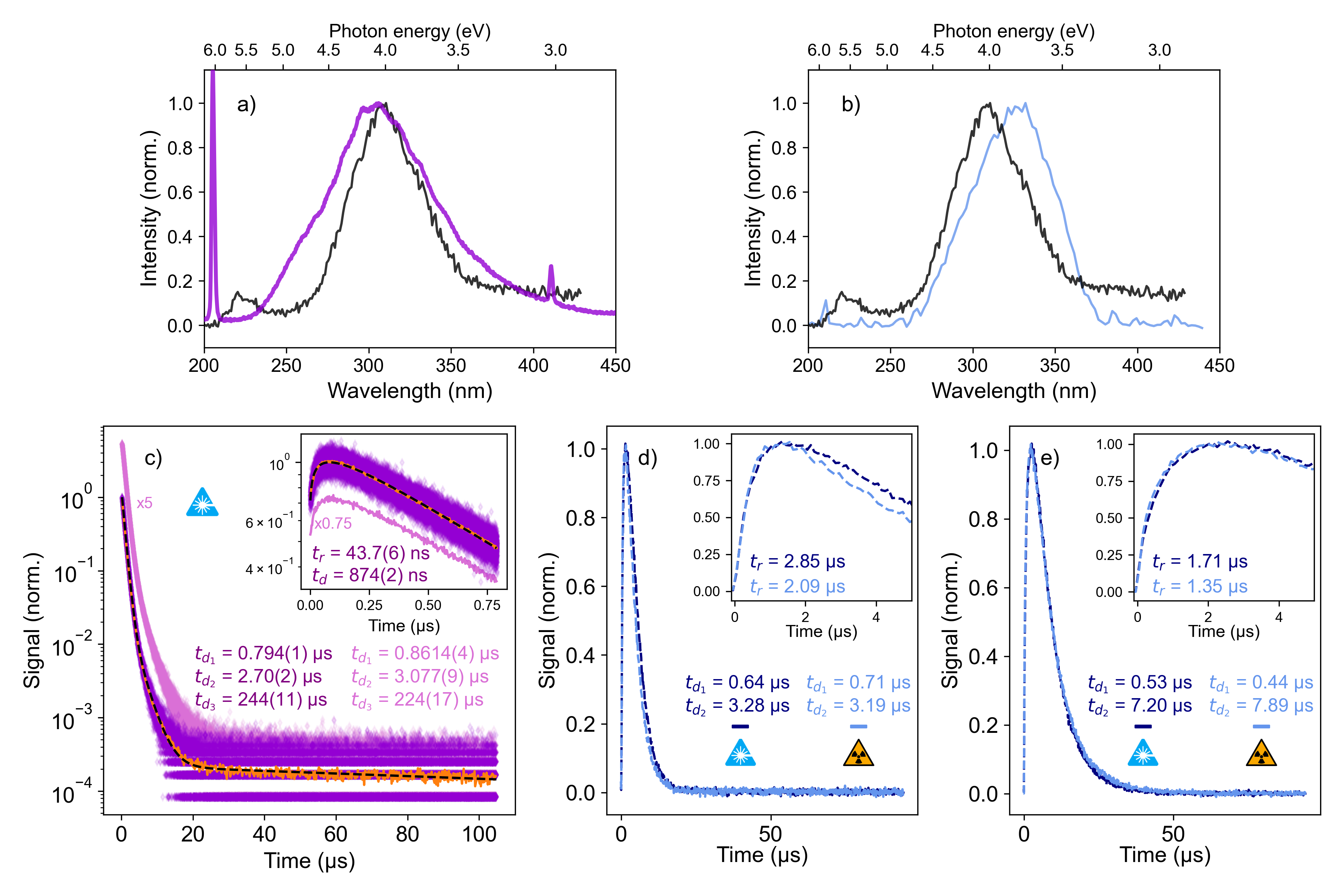}
    \caption{
    \textbf{Spectrally- and temporally-resolved study of scintillation and OSL emission in \ce{BaF2}.}
    \textbf{a)} Spectrally resolved emission of scintillation from the Hellma crystal measured under electron irradiation from a $^{90}$Sr/$^{90}$Y source (black) and two-photon absorption (2PA) excitation with \SI{205}{\nano\meter} \SI{100}{\femto\second}-pulses (purple). The sharp peaks (\SI{205}{\nano\meter} and \SI{410}{\nano\meter}) in the 2PA spectra correspond to stray laser light.
    \textbf{b)} Spectrally resolved emission of scintillation during electron irradiation (black), same as in panel a), and OSL (blue) for the optical-grade crystal.
    \textbf{c)} Time-resolved decay curve of photoluminescence under two photon excitation with 205 nm photons measured on the Hellma (purple) or Epic (light purple) crystals. The orange line corresponds to a binning of 50 points. The inset shows a separate measurement finding the rise time with a finer resolution. The data has been normalized. For the Epic crystal only binned data is shown and it has been scaled by a factor 5 or 0.75 for visual purposes. To remove stray light from the excitation laser, the first \SI{7.5}{\nano\second} and \SI{180}{\nano\second} were discarded for the short and long measurements, respectively.
    \textbf{d)}-\textbf{e)} Time-resolved OSL from the Hellma (d) and Epic (e) samples after pulsed stimulation (wavelength 532 nm) of trapped carriers, which were generated by fs-laser excitation (dark blue) or electron irradiation (light blue). The insets are early-time views of the same data. The data has been binned corresponding to 50 points, and the raw Hellma data can be seen in Supplementary S3.
    }
    \label{fig:3}
\end{figure*}

The observed scintillation spectrum of \ce{BaF2}, shown in Fig.~\ref{fig:3}a-b (black line), is dominated by the STE contribution around \SI{300}{\nano\meter} with a weaker cross-luminescence component at \SI{220}{\nano\meter}\cite{Dorenbos1993}. We find that the same, slightly broadened, STE spectrum is observed during optical excitation using a two-photon absorption (2PA) process to excite electrons across the \SI{10}{\electronvolt} bandgap 
(purple curve in Fig.~\ref{fig:3}a). Electron irradiation and subsequent optical stimulation was performed to acquire the OSL emission spectrum (blue curve in Fig.~\ref{fig:3}b),  showing a slightly redshifted spectrum compared to the STE$_\text{scint}$. The similarities between the spectra indicate that the STE is formed in all instances.
The spectral analysis is additionally discussed in Supplementary S2.

Further, we measure the decay kinetics of the STE emission in two different ways. First, 2PA is used to acquire the decay curves from the STE luminescence shown in 
Fig.~\ref{fig:3}c for both samples. These curves are fitted to three decay components, 
and the resulting values are presented in the figure, showing similar decay components for both samples, considering the 15\% variability found in literature. The main decay of the STE$_\text{scint}$ created by 2PA agrees with previously reported values of \SI{600} to \SI{800}{\nano\second} \cite{Pots2020,Lindner2001,Demidenko2010}. The two slower decays of the STE$_\text{scint}$ of around \SI{3}{\micro\second} and \SI{200}{\micro\second} have also been reported previously \cite{Lindner2001}. 
The figure inset shows a measurement with finer resolution at short times, revealing a rising edge and is thus fitted to a single rise and decay model. In combination with the spectrally-resolved measurements, this study allows us to characterize the STE$_\text{scint}$, which appears to have the same spectral and decay properties in both crystals. 

After exposure to either electron irradiation or fs-laser pulses, the samples were transferred to the setup sketched in Fig.~\ref{fig:2}, which measures the STE$_\text{OSL}$ decay kinetics upon pulsed optical stimulation of trapped charge carriers. The decay curves are presented in Fig.~\ref{fig:3}d-e showing next to identical kinetics when measuring the same crystal under different irradiation conditions. Fitting to a model with one rise and two decay components yields the fitting values shown in the figure, where the first and second decay components of the STE$_\text{scint}$ emission are recovered for the optical-grade sample. Analyses with a one decay component model 
are compared in Supplementary S3.
The discrepancy seen between the decay components of the STE$_\text{scint}$ and STE$_\text{OSL}$ in the scintillator-grade sample, points to an STE$_\text{OSL}$ that is not composed by the initial \emph{F}~+~\emph{H} centers, as shown in Fig.~\ref{fig:1}d-e. We associate these differences with impurities present in this sample which act as trapping sites for the STH. 
As confirmed by this study, the decay kinetics are independent of the initial irradiation source, pointing to the reformation of the STE upon optical stimulation.

\subsection{Optical absorption bands and OSL yield manipulation}
\begin{figure*}[h]
    \centering
    \includegraphics[width=0.9\linewidth]{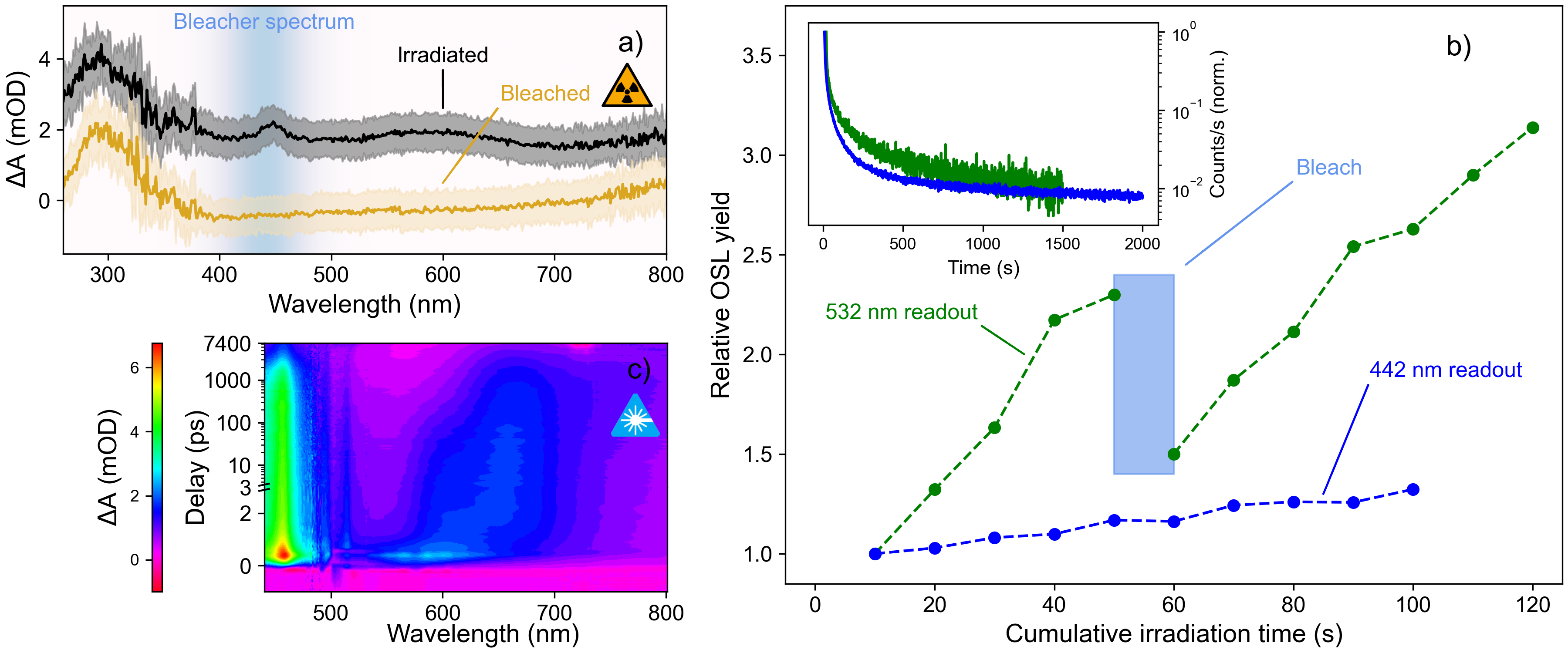}
    \caption{
    \textbf{Investigation of optical absorption and increasing OSL yield.}
    \textbf{a)} Absorption spectrum of \ce{BaF2} after two hours of electron irradiation (black) and after subsequent optical bleaching (yellow). Respective standard deviations are shown as shaded bands.
    The vertical blue shading indicates the spectrum of the bleaching lamps.
    \textbf{b)} Intensity of the integrated OSL signal after successive irradiations and readouts relative to the intensity after the first irradiation step. The blue box represents an optical bleach with the same lamps shown in (a). The inset shows an example of an OSL curve, stimulating with either \SI{532}{\nano\meter} (green) or \SI{442}{\nano\meter} (blue).
    \textbf{c)} Transient absorption data measured after pulsed excitation at \SI{240}{\nano\meter}. The delay represents the time difference between the pump and probe laser pulses. Colors represent optical absorption as indicated by the color bar. Note the change to log scale after \SI{3}{\pico\second}. The data is a combination of two separate measurements, switching from one to the other at \SI{500}{\nano\meter} due to noise.
    }
    \label{fig:4}
\end{figure*}

The steady-state optical absorption spectrum of \ce{BaF2} following 
electron irradiation is shown in Fig.~\ref{fig:4}a (black curve). The absorption is measured relative to a reference spectrum taken before irradiation. Three distinct absorption bands are observed, two in the visible region, peaking at around \SI{450}{\nano\meter} and \SI{600}{\nano\meter}, and one in the UV, centered around \SI{300}{\nano\meter}. The crystal is subsequently bleached using lamps with a spectrum indicated by the shaded blue region in Fig.~\ref{fig:4}a, and its optical absorption spectrum is remeasured and shown in yellow in the same figure. Based on the disappearance of the visible bands after optical stimulation, we propose that these two bands must correspond to the electron and hole traps responsible for the OSL signal. Further, literature values for the optical absorption of \emph{F}~centers \cite{Rodnyi2024, Williams1976, Nepomnyashchikh2002} are in good agreement with the observed \SI{600}{\nano\meter} band, and we thus propose that the \SI{450}{\nano\meter} band corresponds to the STH. The band centered at \SI{300}{\nano\meter} remains unaffected and therefore cannot be the primary contributor to OSL.

To validate our band designations, we monitor the depletion rate of the OSL signal using continuous-wave optical stimulation, exemplified by the decay curves shown in the inset of  Fig.~\ref{fig:4}b. The integral of this curve is proportional to the OSL yield, i.e., the number of OSL photons emitted per given amount of energy deposited \cite{Mads2022}. Repeated measurements of this kind allow us to verify whether the crystal returns to initial conditions after irradiation and subsequent optical stimulation. 
We choose to stimulate with \SI{442}{\nano\meter} photons as this overlaps with the \SI{450}{\nano\meter} optical absorption band that we propose will stimulate the STH via process 6 in Fig.~\ref{fig:1}a. Similarly, we choose \SI{532}{\nano\meter}, which approaches the \emph{F}~band that we propose will stimulate process 7 in Fig.~\ref{fig:1}a. 
Notice that in the latter process, the electron moves through the conduction band and recombines with its trapped hole counterpart, leaving behind a fluor vacancy and a \ce{F-} interstitial ion (Frenkel pair).
In contrast, the separated defects shown in Fig.~\ref{fig:1}f will only be capable of restoring the initial configuration, i.e., with the fluorine interstitial returning to its vacancy position, when stimulating the STH. Hence, we expect STH stimulation to recreate the initial condition to a higher degree than \emph{F} center stimulation. 
Indeed, as observed in Fig.~\ref{fig:4}b, \ce{BaF2} shows a relatively constant OSL yield after repeated but identical irradiations lasting \SI{10}{\second} when using \SI{442}{\nano\meter} blue stimulation, while this yield increases when stimulating with \SI{532}{\nano\meter} green photons. 
Notice also that the inserted blue-light bleaching sequence counteracts the OSL increase observed with green light readout. Altogether, the above observations strongly support our designations of the STH and \emph{F} center bands,
with the \SI{450} and \SI{600}{\nano\meter} bands corresponding to the hole and electron part of the STE$_\text{OSL}$, respectively.
The creation of the proposed \emph{F} and \emph{H} bands, associated with long-lived electron–hole pairs responsible for the OSL signal, was further investigated after 2PA with a sub-\SI{}{\pico\second}, transient-absorption setup. Fig.~\ref{fig:4}c shows the two bands, now centered at \SI{456}{\nano\meter} and around \SI{650}{\nano\meter}, although the latter peak position is obscured by an experimental artifact around \SI{720}{\nano\meter}. 
We see that the bands form immediately after excitation, indicating a clear relation to the inherent scintillation process sketched in Fig.~\ref{fig:1}b-e. 
An additional short-lived band is found around \SI{580}{\nano\meter}. This band decays quickly with a decay time of \SI{700}{\femto\second}, correlated with the rise time of the \SI{650}{\nano\meter} band. The same decay is seen for the band at \SI{456}{\nano\meter}, followed by a slower main decay around \SI{5}{\nano\second}. The band at \SI{650}{\nano\meter} has a decay time of around \SI{7}{\nano\second}. Further details can be found in Supplementary S4.

\subsection{Spatial mapping of the OSL signal}

\begin{figure*}[ht]
    \centering
    \includegraphics[width=0.8\linewidth]{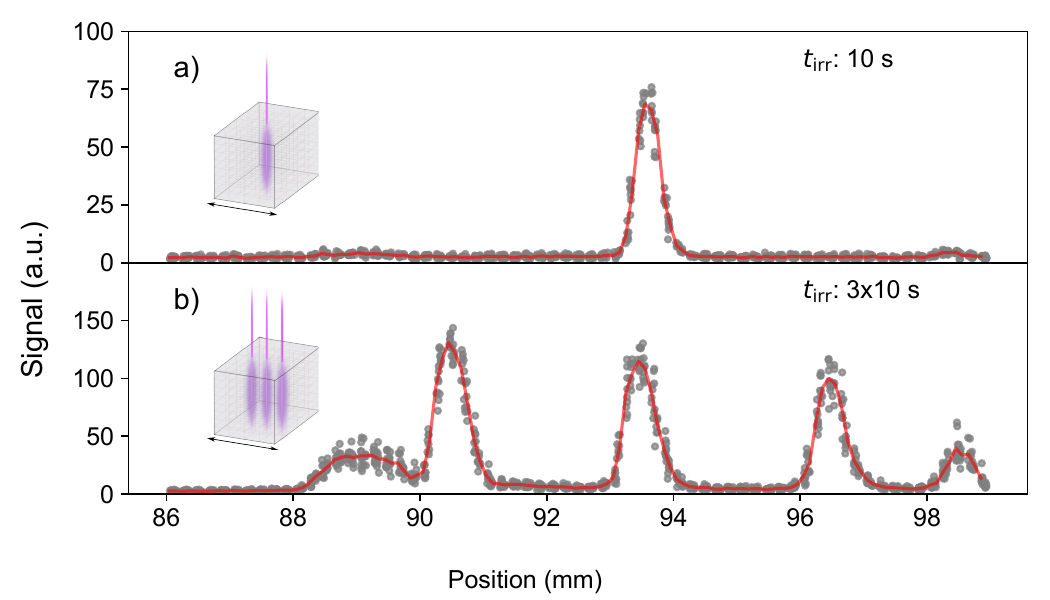}
    \caption{
    \textbf{Spatially-resolved imaging of fs-laser-induced charge trapping.} The beam widths of the \SI{205}{\nano\meter} laser beam at the crystal surface is around \SI{30}{\micro\meter}. The red lines show running averages.
    \textbf{a)} Focusing the laser to one spot.
    \textbf{b)} Focusing the laser to three spots separated by 3 mm. 
    The additional peaks on either side are a result of light scattering at the crystal edges interacting with the trapped charges and giving an unwanted readout.
    The width is limited by the readout laser.}
    \label{fig:5}
\end{figure*}

The setup in Fig.~\ref{fig:2} can spatially map the OSL signals. Such results are shown in Fig.~\ref{fig:5}, where OSL is read out along one axis of the crystal by translating the stimulating laser. The readout was preceded by fs-laser excitation (\SI{205}{\nano\meter} wavelength) orthogonal to the translation axis at either one specific position (a) or three positions placed in \SI{3}{\milli\meter} increments (b). 
The spatial resolution of the setup is limited by the width of the around \SI{500}{\micro\meter}-wide readout laser.
These results showcase how OSL-active traps can be created and placed by optical excitation.



\section{Discussion}
When examining the scintillation spectra, from both electron irradiation and 2PA, and the OSL spectrum in Fig.~\ref{fig:3}a–b, we observe clear similarities alongside subtle differences in the emission. The 2PA spectrum is consistent with previous studies that attribute multiple spectral components to distinct lattice configurations of the STE \cite{Lindner2001}. That study \cite{Lindner2001}  shows how the STE$_\text{scint}$ emission redshifts at time intervals beyond \SI{4}{\micro\second}. This aligns well with the redshift observed for the STE$_\text{OSL}$ signal, which  decays with a \SI{3} to \SI{7}{\micro\second} decay component depending on the crystal. The specific lattice configurations of the STE$_\text{OSL}$ for a given stimulation wavelength remains to be determined.

The STE$_\text{scint}$ luminescence measured under two-photon excitation in Fig.~\ref{fig:3}c does not rise instantaneously, but instead has a rise time around \SI{45}{\nano\second}. This has also been reported by others and is theorized to be related to a population of long-lived traps feeding into the STE population, which would depend on the excitation density of initial electron-hole pairs\cite{Fedorov2009, Kimura1998}. 
Assuming this explains the rise, the slower rise observed for the OSL signal could be a signature of a lower population density of long-lived traps, as only a small fraction of the STE$_\text{scint}$ result in trapped charges. This causes a larger separation of trapped carriers, leading to the observed OSL signal rising slowly, while decaying similarly to the STE$_\text{scint}$.

Our assignment of the \emph{H}~band in Fig.~\ref{fig:4}a,c centered at \SI{450}{\nano\meter} challenges previous measurements with Tm-doped \ce{BaF2} samples proposing the band to be around \SI{330}{\nano\meter}\cite{Beaumont1970}. Experimental conditions, especially temperature and sample purity, can affect the absorption spectra immensely, as attempts to reproduce the results have conversely shown the band to be absent in pure samples and instead find a different band attributed to \emph{V$_K$} centers around \SI{365}{\nano\meter}\cite{Nepomnyashchikh2002, Cavenett1969, Williams1976}. The authors argue that the \emph{H}~band has been assigned erroneously and is rather due to the Tm-doping. 
Notice that all referenced results in literature are measured at or below \SI{77}{\kelvin}, and temperature plays an important role, 
especially as the dominant \emph{V$_K$}~centers become mobile only above \SI{110}{\kelvin}\cite{Call1974, Popov2017}. Additionally, DFT simulations have suggested a transition energy of the \emph{H}~center corresponding to \SI{420}{\nano\meter}\cite{Jia2010}, remarkably close to our band
.
We believe that we have identified the \emph{H}~band at room temperature at \SI{450}{\nano\meter} from steady-state and transient absorption. This is further corroborated by the relative OSL yields shown in Fig.~\ref{fig:4}b, which do not increase when stimulating the \emph{H} band. Conversely, the OSL yield increases when stimulating only the trapped electrons as this mechanism (process 7 in Fig.~\ref{fig:1}a) is unable to recover the perfect lattice, leaving behind vacancy-interstitial pairs that can act as direct traps for electrons and holes in subsequent irradiations.

The transient absorption data in Fig.~\ref{fig:4}c makes clear that physical charges are created directly and immediately by excitation across the bandgap. The relaxation to stable defects can also be observed, where the absorption from \emph{H}~centers is immediately present at \SI{456}{\nano\meter}, while electrons first absorb around \SI{580}{\nano\meter} before being trapped at the vacancy left from the rotation of the \emph{H}~center. 
This rotation has also been suggested by simulations\cite{Chuklina2019}.
Our interpretation arises from the kinetics of the absorption showing the same \SI{700}{\femto\second} decay at both \SI{456} and \SI{580}{\nano\meter}, while the \emph{F}~band above \SI{600}{\nano\meter} rises within the same time scale. We propose that \SI{700}{\femto\second} is the time it takes for the \emph{H}~center to form a fluor interstitial, giving rise to a fluor vacancy where the electron localizes.  

Further investigations in the UV region of the spectrum are needed to identify the nature of the stable UV absorption band, which appears slightly blue-shifted compared to previously reported values for transient \emph{V$_K$}~centers\cite{Williams1977}. The absence of transient absorption data at these wavelengths hinders a complete understanding and is therefore a focus for future work. We also propose measuring the wavelength-dependent cross section of the OSL process\cite{Warming2026} to improve insight into the initial and final states of the trapped electrons and holes.

\subsection{Conclusion}
The results presented here provide a purely optical method for studying the dynamics of trapped carriers in ionic crystals of the fluorite family. 
Further, the presented readout method can be used to spatially map the distribution of any excitation above the bandgap in the crystal, using the OSL traps as an optical memory. This has applications in radiation detection.

To summarize, we have shown that while the scintillation mechanism in \ce{BaF2} forms an STE in the known \emph{F}~+~\emph{H} lattice configuration, the non-radiative decays of such STEs lead to spatially separated \emph{F} and \emph{H}~centers. 
The spatially separated \emph{F} and \emph{H}~centers are the long-lived electron- and hole-traps responsible for the OSL signal. 
The STE is similarly responsible for the photon emission in the OSL process, however it can also form an \ce{e-}~+~\emph{H} structure.
The stimulation wavelength can be chosen to preferentially interact with one trap over the other, and the release of electrons from \emph{F}~centers leads to an increase in the OSL yield in the crystal upon subsequent irradiations.
Finally, we show that spatial information of the OSL signal from trap formation can be recovered using a scanning laser sheet. These results highlight that optical excitation creates long-lived vacancy-interstitial pairs which are otherwise usually regarded as a result of radiation damage from ionizing radiation. Our interpretation challenges recent investigations in \ce{LiF} that attribute \emph{F}~center formation to direct momentum transfer from the ionizing particle, leading to the displacement of an entire fluorine atom and the creation of stable defects\cite{Pop2025, Massillon2018}.
Instead, we show here that the formation of vacancy-interstitial pairs is intrinsically linked to the creation of self-trapped holes and can be triggered by any excitation above the bandgap.

\section{Methods}

\subsection{Setup for spatially- and temporally-resolved optically stimulated luminescence}
Our sensitive setup for OSL measurements shown in Fig.~\ref{fig:2} utilizes a custom-built high-frequency circuit and the single-photon sensitivity of silicon photomultipliers (SiPMs) to measure the individual time stamps of hundreds of photons after pulsed stimulation. The board also outputs a regular integrated SiPM signal\cite{Gundacker2019, Krake2022}.
The stimulation source is a \SI{532}{\nano\meter} \ce{Nd{:}YAG} laser operating at \SI{10}{\hertz}, and a collimated laser sheet with a width of about \SI{500}{\micro\meter} is produced by two cylindrical lenses with focal lengths of \SI{1000}{\milli\meter} and \SI{-100}{\milli\meter}, respectively. The height of the beam is set with an iris to be slightly less than the crystal height to avoid scattering from the edges.

The setup can operate in two modes, either stationary acquisition of decay curves or spatially mapping the OSL signal along one axis.
The decay kinetics are measured by combining photon arrival times from multiple pulses. This can produce a decay curve using fewer pulses than conventional time-correlated single-photon counting, which is useful when measuring OSL where the signal decreases with cumulative stimulation energy. 
In order to carry out spatially resolved measurements, the setup contains a DDS100 translation stage from Thorlabs which moves two mirrors, translating the sheet along the crystal. 
Common to both modes is the imaging of the front surface of the crystal onto a 3x3 \SI{}{\milli\meter\squared} AFBR-S4N33C013 SiPM from Broadcom. The imaging is done with two aspheric UV lenses (ASL2520-UV) from Thorlabs on either side of two Thorlabs FGUV11-UV filters mounted to reduce the amount of scattered laser light. For each pulse, the measured signal is acquired on a Lecroy WavePro 404HD oscilloscope, running a Python program that also controls the laser firing and stage positioning. 
The collection efficiency of the setup depends on the laser position relative to the lens because of the increasing solid angle through the crystal becoming larger than the detector sensor area. This is apparent with the weakening signal of the peaks in Fig.~\ref{fig:5}b, despite equal irradiation times. 

\subsection{Spectrally-resolved measurements}
To measure the emission spectrum of radioluminescence from \ce{BaF2}, an optical fiber was guided from an irradiation chamber with a $^{90}$Sr/$^{90}$Y electron source into a Princeton Instruments Acton SP2300
spectrometer coupled to a Pixis-100 CCD.
Similarly, OSL spectra were measured with the same spectrometer using a \SI{460}{\nano\meter} laser diode for stimulation. This required the addition of optical filters ThorLabs FGUV11 and FGUV5 in front of the  spectrometer, providing a transmission window between \SI{270} and \SI{370}{\nano\meter}. The upper limit of this window is near the highest wavelength from the \ce{BaF2} emission.
Lastly, emission spectra following two-photon absorption (2PA) were acquired using a fiber-coupled Lasertack LR2 spectrometer while exciting the crystal with focused \SI{205}{\nano\meter} \SI{100}{\femto\second}-pulses.

\subsection{Setup for pulsed laser excitation}
The \ce{BaF2} crystals were excited by 2PA using the fourth harmonic of a \SI{5}{\kilo\hertz} Ti:sapphire pulsed-laser system with \si{\femto\second} pulse duration. The setup utilizes second-harmonic generation in two $\beta$-Barium borate (BBO) crystals to change the photon energy/wavelength from the fundamental \SI{1.51}{\electronvolt}/\SI{820}{\nano\meter} to \SI{3.02}{\electronvolt}/\SI{410}{\nano\meter} and finally to \SI{6.04}{\electronvolt}/\SI{205}{\nano\meter}. The generated fourth-harmonic pulses are then focused to a beam width of around \SI{20}{\micro\meter} at the beam waist by a UV-grade fused silica lens. Both the crystal holder and the lens are placed on translation stages which allows us to move the beam waist to any position within the sample.

Time-resolved measurements of the laser-excited \ce{BaF2} crystals are done with time-correlated single-photon counting using a PicoQuant 07 Hybrid PMT mounted with two Thorlabs FGUV11-UV bandpass filters. The PMT has a Gaussian impulse response function with a full width at half maximum of around \SI{70}{\pico\second}.

The above-mentioned setup was also used to laser-excite the samples prior to OSL measurements.
The beam width at the beam waist was around \SI{20}{\micro\meter}.
The location of the irradiated region in the crystal was controlled by moving the crystal with a translation stage, irradiating at either one position or three positions in increments of \SI{3}{\milli\meter} with a micrometer screw. The irradiation time was around \SI{10}{\second} at \SI{5}{\kilo\hertz} with pulse energies of \SI{1}{\nano\joule}.

\subsection{Absorption}
\subsubsection{Steady-state absorption}
Experiments of stable absorption were done upon electron-irradiation of a \SI{1}{\cubic\centi\meter} cubic Hellma \ce{BaF2} crystal. The irradiation source was a $^{90}$Sr/$^{90}$Y sample with an activity of around 64 MBq.
The crystal was first measured to get a reference spectrum, followed by a measurement after two hours of irradiation. Finally, the crystal was bleached optically for \SI{30}{\minute} using two arrays of blue LEDs, giving an intensity of about \SI{25}{\milli\watt\per\square\centi\meter}, before the absorption spectrum of the bleached crystal was acquired.
The reference spectrum is used as a baseline for the absorption of the crystal. This is subtracted from the irradiated and bleached spectra, so we can ensure that these spectra only contain absorption relating to the given electron irradiation and not the possible prior irradiation history of the sample.
The absorption was measured nine times for each sample using a High-Performance Lambda Spectrometer from PerkinElmer, and the mean and standard deviations of those measurements were calculated.

\subsubsection{Transient absorption}
The measurements of transient absorption data utilized the HARPIA-TA system from Light Conversion. The \SI{240}{\nano\meter} \SI{}{\femto\second} excitation pulses are produced by an OrpheusNeo optical parametric amplifier and harmonic generator also from Light Conversion. The probe pulses are produced by white-light supercontinuum generation. The time between excitation and probe pulses is controlled by a delay stage capable of covering a delay of \SI{8}{\nano\second}. The sample was a 3x3x1\SI{}{\milli\meter\cubed} \ce{BaF2} crystal. Two-photon absorption at \SI{240}{\nano\meter} is sufficient to bridge the bandgap of \ce{BaF2}. Due to dispersion, the white light supercontinum pulses are chirped. Therefore, the absorption will be delayed by varying amounts across the spectrum, so a chirp correction had to be made. This was done by finding the rising edge of the absorption by the steepest gradient. After smoothing, the data was then fitted as a function of wavelength to the following expression\cite{Crane2023}
\begin{equation}
 t_0(\lambda) = A + \frac{B}{\lambda} + \frac{C}{\lambda^2},
\end{equation}
where $\lambda$ is the wavelength and A, B, and C are fitting parameters for the constant, linear, and quadratic inverse wavelength terms, respectively. 
A notch filter was inserted to block the fundamental wavelength at 720 nm in order not to saturate the spectrometer.

\subsection{OSL yield determination}
The OSL yield was measured with a setup consisting of a light tight box with an optical window allowing the introduction of continuous-wave laser-diode light at \SI{532}{\nano\meter} or \SI{442}{\nano\meter}. 
In the box, the light is collected using a lens with a focal length of \SI{10}{\centi\meter} onto a photomultiplier tube (PMT) mounted with two FGUV11 filters to separate the OSL photons from the stimulation light. The PMT was set to count single photons in \SI{1}{\second} intervals, and the readouts lasted for \SI{1500}{\second} and \SI{2000}{\second} for the readout wavelengths of \SI{532}{\nano\meter} and \SI{442}{\nano\meter}, respectively. 



\bibliography{Bibliography_standardized}

\end{document}